\documentclass[fleqn,twoside,twocolumn,nofootinbib,showkeys]{revtex4}
\usepackage[sec,nocpr]{ujp}
\usepackage[utf8]{inputenc}
\usepackage{graphicx}
\usepackage[colorlinks,hypertexnames=false]{hyperref}
\razd{\seci}
\usepackage{myphi}

\newcommand{\sNNn}[1]{\sNN $=$ \SI{#1}{\GeV}}

\graphicspath{{picts/}}

\begin{document}
\title[New results on \phiM
production from the \NASixtyOne experiment at CERN SPS]{NEW RESULTS ON
$\pmb{\phi}$(1020) PRODUCTION FROM THE \NASixtyOne EXPERIMENT AT CERN SPS}
\author{A.~Marcinek for the \NASixtyOne Collaboration}
\affiliation{Institute of Nuclear Physics, Polish Academy of Sciences}
\address{ul. Radzikowskiego 152, PL-31342 Kraków, Poland}
\email{antoni.marcinek@ifj.edu.pl}

\razd{\seci}

\autorcol{A.~Marcinek for the \NASixtyOne Collaboration}



\begin{abstract}
\NASixtyOne is a multipurpose, fixed-target hadron spectrometer at the CERN SPS.
Its research program includes studies of strong interactions as well as
reference measurements for neutrino and cosmic-ray physics. A significant
advantage of \NASixtyOne over collider experiments is its extended coverage of
phase space available for particle production. This includes the entire
projectile hemisphere of the collision, with no low-\pt cut-off.
\par
The energy and system-size dependence of strangeness production plays an
essential role in studies of the transition from confined to deconfined matter.
With its zero net strangeness and its valence structure composed predominantly
of $s$ and $\bar{s}$ valence quarks, the \phiM meson will not be sensitive to
strangeness-related effects in a purely hadronic scenario, but will behave like
a doubly strange particle in a partonic system.
\par
This contribution presents the first-ever results on \phiM meson production in
intermediate-size systems at the CERN SPS, that is, central \ArSc
collisions at \mbox{$\sqrt{s_\mathrm{NN}}$ = 8.8}, 11.9, and 16.8 GeV. The
presented results include double-differential rapidity-transverse momentum
(\y-\pt) distributions,
transverse mass (\mt) spectra at midrapidity, \pt-integrated rapidity spectra, mean
multiplicities ($4\pi$ yields), and particle ratios. These are compared to data
on \PbPb and \pp collisions. A discussion of open and hidden strangeness
production enhancement is included. Finally, a comparison with three microscopic
models is shown, demonstrating their overall failure in describing these new
measurements.
\end{abstract}


\maketitle

\section{Introduction}
\begin{figure*}[t]
  \centering
  \includegraphics[width=0.8\textwidth]{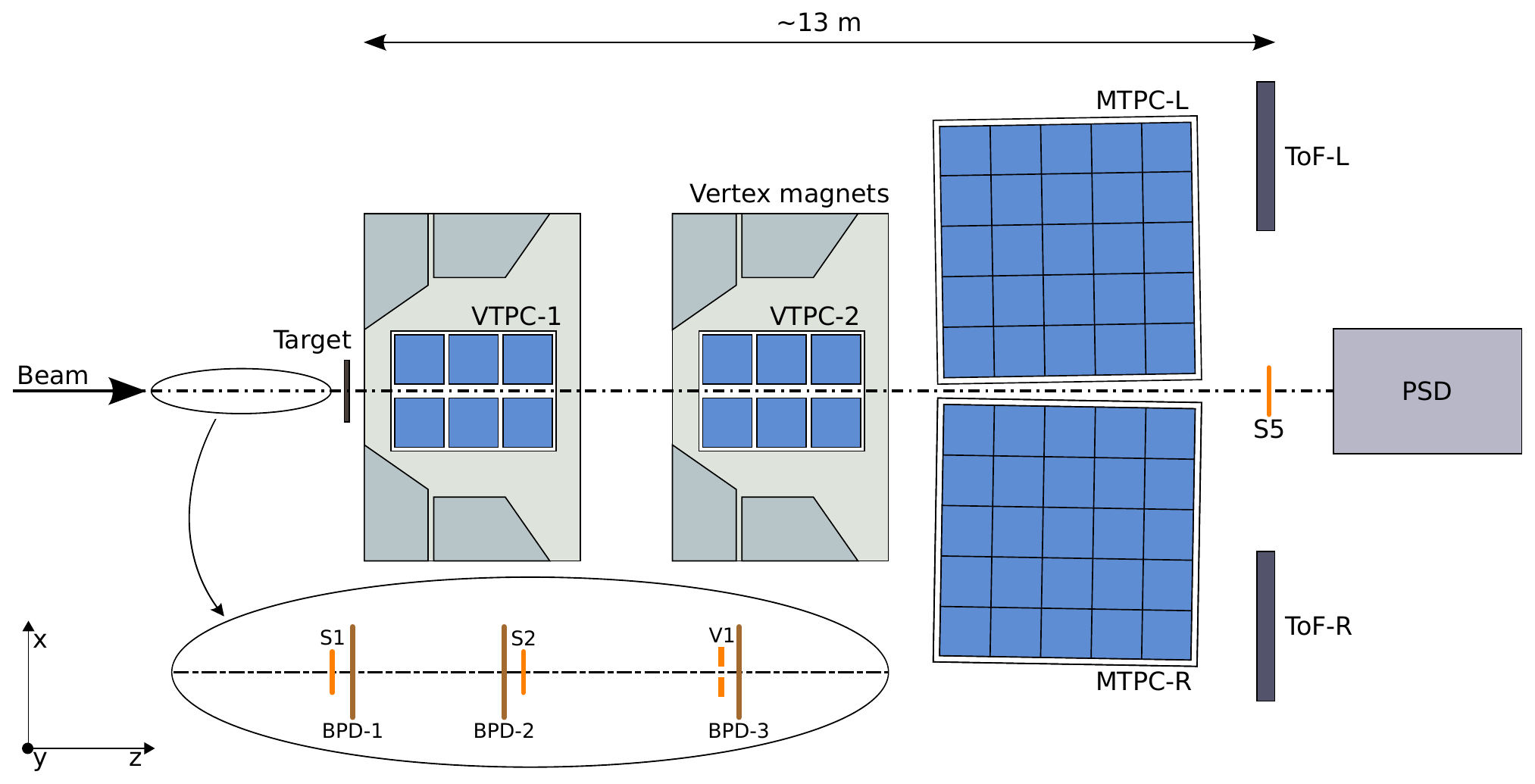}
  \caption{Schematic layout of the \NASixtyOne detector system (horizontal cut
    in the beam plane, not to scale) showing the state of the detector during
    \ArSc data taking in 2015.
  }
  \label{fig:det}
\end{figure*}
The subject of this contribution are preliminary results on \phiM production in
central \ArSc collisions measured by the \NASixtyOne experiment at the CERN SPS,
obtained using the $\phi \to \Kp\Km$ decay channel.
Associated with the idea that the energy and system-size dependence of
strangeness production plays an essential role in studies of the transition from
confined to deconfined matter~\cite{Gazdzicki:2010iv}, this work has a two-fold
motivation. First, with its zero net strangeness and its valence structure
composed predominantly of $s$ and $\bar{s}$ valence quarks, the \phiM meson is
of particular interest for constraining hadron production models: \phiM will not
be sensitive to strangeness-related effects in a purely hadronic scenario, but
will behave like a doubly strange particle in a partonic system. Second, \ArSc
holds an intermediate position between earlier-measured minimum bias
\pp~\cite{NA61SHINE:2019gqe, NA49:2000jee} and central \PbPb~\cite{NA49:2008goy}
collisions, thus allowing for an improvement of our knowledge of \phiM-related
phenomena as a function of system size.
\par
The \NASixtyOne detector~\cite{bib:NA61_facility} is a multipurpose fixed-target
spectrometer at the CERN SPS. Over the years the detector underwent multiple
upgrades.
\Cref{fig:det} shows its state during the \ArSc data taking in 2015.
Its main components are
large-volume Time Projection Chambers (TPCs), two of them (VTPCs) immersed in a magnetic
field perpendicular to the beam. This gives \NASixtyOne a significant advantage
over collider experiments --- acceptance covering nearly entire forward c.m.s.\
hemisphere, down to $\pt=0$ for charged hadrons. In addition to providing
momentum measurement, TPCs provide also particle identification (PID)
capabilities using energy loss (\dEdx) of charged particles in the gas, which is
further augmented with time-of-flight measurements in limited phase-space range.
Centrality is measured via forward energy deposit in a hadronic calorimeter, the
Projectile Spectator Detector (PSD).

\section{Analysis methodology}
\begin{figure}[t]
  \centering
  \includegraphics[width=0.5\textwidth,page=22,clip,trim=35mm 0 35mm 0]{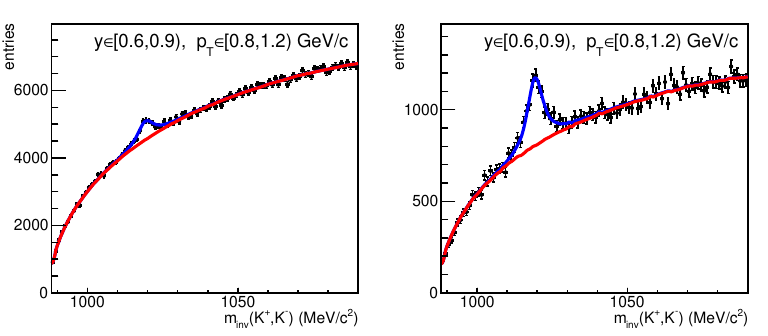}
  \caption{Illustration of the tag-and-probe method (see text) on one of
    analysis bins for \ArSc collisions at \sNNn{16.8}. Left: invariant mass
    spectrum for the tag sample where at least one of kaon candidates in the
    pair needs to pass the strict PID cut. Right: invariant mass spectrum for
    the probe sample where both kaon candidates in the pair need to pass the
    strict PID cut. Red curves are fitted background contributions, while blue
    curves show the sum of background and signal components.
  }
  \label{fig:tp}
\end{figure}
\begin{figure*}[t]
  \centering
  \includegraphics[width=0.8\textwidth,page=2]{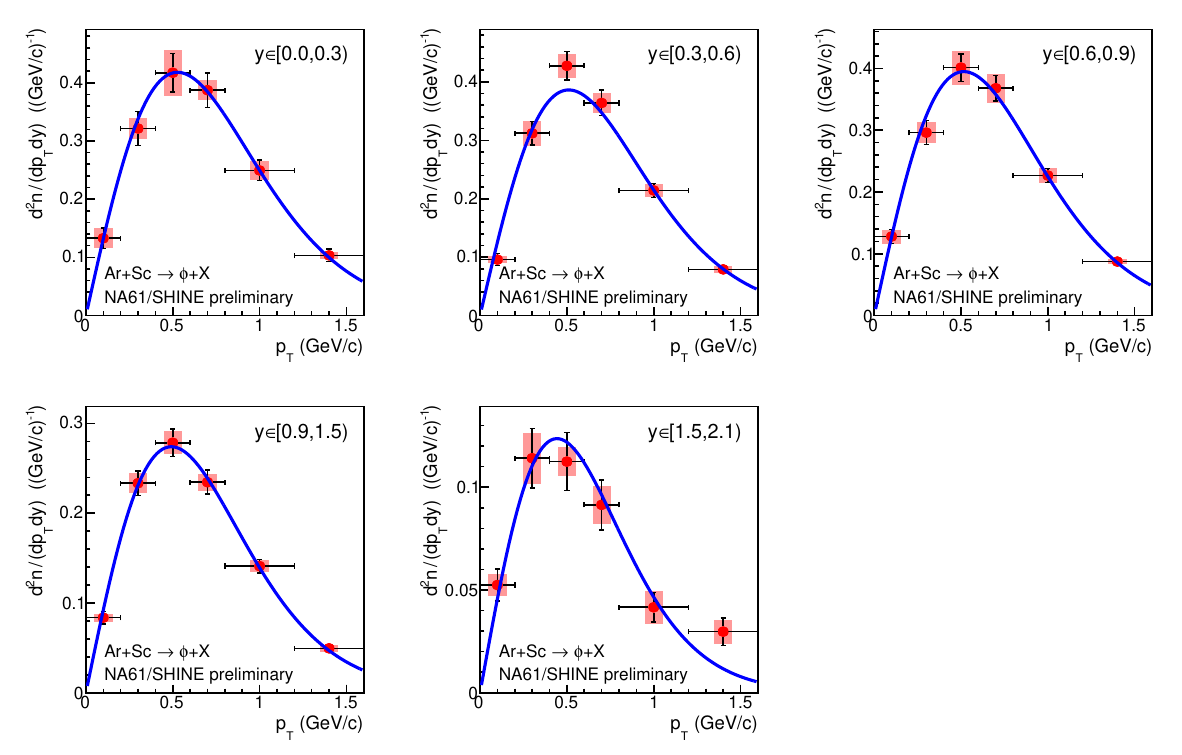}
  \caption{
    Double-differential (\y, \pt ) spectra of \phiM produced in the
    10\% most central \ArSc collisions at \sNNn{16.8}.
    Vertical bars represent statistical uncertainties, red rectangles systematic
    uncertainties, and horizontal bars depict \pt bin sizes.
    Blue curves are fits used to obtain the integrals of the unmeasured tails of
    the \pt spectra to calculate \pt-integrated $\dv*{n}{\y}$ distribution.
  }
  \label{fig:pt}
\end{figure*}
The first step of the analysis is selection of data. In case of \phiM analysis
in \ArSc collisions, 10\% most central events were chosen using forward energy
deposited in PSD, without pileup, with well measured main vertex, and occurring
in the target. Given that \phiM mesons, from experimental perspective, decay
within the main vertex, well reconstructed tracks coming from the main vertex
were selected for the analysis, with enough points in TPCs to assure accurate
momentum and \dEdx measurement.
\par
Special attention was given to particle identification selecting tracks whose
\dEdx was inside bands around kaon Bethe-Bloch curve. In the old
analyses~\cite{NA49:2000jee,NA49:2008goy} a single sample of kaon candidates was
chosen with narrow \dEdx cuts to suppress non-kaon background. Because these
cuts removed also significant amount of kaons, they also removed \phiM mesons. This
bias was corrected for assuming precise knowledge of \dEdx distributions for
kaons and therefore came with significant systematic uncertainty. In \NASixtyOne
a different approach was chosen~\cite{NA61SHINE:2019gqe}, using the
\emph{tag-and-probe} method~\cite{ATLAS:2014pju, LHCb:2011ijs}. First, for every
track an outer band of $\pm 13\%$ of kaon Bethe-Bloch value was used to suppress
non-kaons, without removing any kaons and thus introducing no bias. Second, the
actual tag-and-probe method was applied utilizing an inner, biasing, band of
$\pm 5\%$ of kaon Bethe-Bloch value. Two samples of oppositely-charged kaon
candidate pairs were created: in the \emph{tag} sample at least one of kaon
candidates in the pair needed to pass the narrow \dEdx cut, while in the
\emph{probe} sample both candidates had to pass the cut.
\par
The method is illustrated in \cref{fig:tp}. To perform the analysis
differentially, tag and probe invariant mass spectra were created in bins of
pairs' rapidity and transverse momentum. Denoting the unknown efficiency of kaon
selection by the narrow cut as \PIDeff and the yield of \phiM mesons
contributing to the spectra as \Nphi, combinatorics of the problem gives the
following expected signal yields in the two samples:
\begin{equation}
  \Nt = \Nphi \PIDeff (2 - \PIDeff) \hspace{2em} \text{and} \hspace{2em} \Np =
  \Nphi \PIDeff^2\,.
  \label{eq:tp}
\end{equation}
The tag and probe spectra were fitted simultaneously to get \Nphi. For both
samples the model curve was a sum of signal and background contributions, where
signal was a convolution of relativistic Breit-Wigner and
q-Gaussian~\cite{NA61SHINE:2019gqe} and background was parametrised as a sum
of event mixing and \KstarM templates.
\begin{figure*}[t]
  \centering
  \includegraphics[width=0.8\textwidth,clip, trim=14mm 39mm 14mm 14mm]{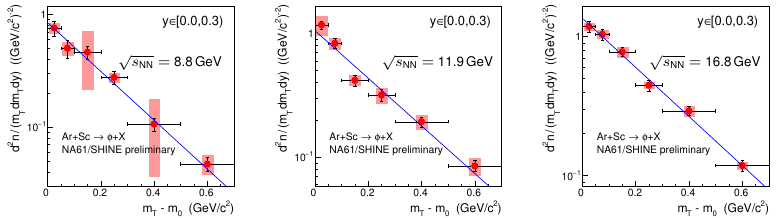}
  \caption{Transverse mass spectra of \phiM mesons produced at midrapidity in
    the 10\% most central \ArSc reactions at three collision
    energies indicated in the plots.
    Vertical bars represent statistical uncertainties, red rectangles systematic
    uncertainties, and horizontal bars depict \pt bin sizes.
    Blue curves are exponential fits to obtain inverse slope parameters (see text).
  }
  \label{fig:mt}
\end{figure*}
\begin{figure*}[t]
  \centering
  \includegraphics[width=\textwidth,clip, trim=9.6mm 31.5mm 9.6mm 13mm]{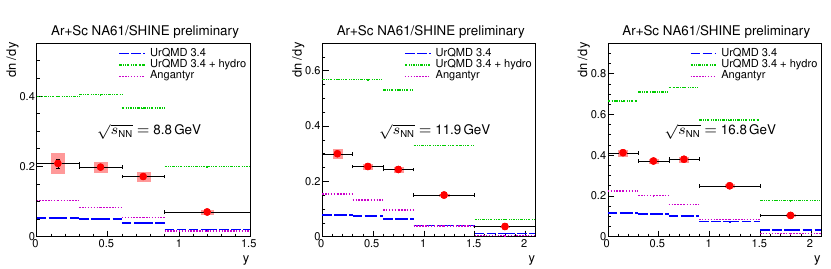}
  \caption{Rapidity distributions of \phiM mesons produced in the 10\% most
    central \ArSc collisions measured by the \NASixtyOne
    experiment (red circles), compared to model predictions.
    Vertical bars represent statistical uncertainties, red rectangles systematic
    uncertainties, and horizontal bars depict \pt bin sizes.
    Model calculations were performed by S.~Veli (Technical University of Munich) and
    T.~Janiec (The University of Manchester).
  }
  \label{fig:dndy}
\end{figure*}
\begin{figure*}[t]
  \centering
  \includegraphics[width=0.3\textwidth, page = 1]{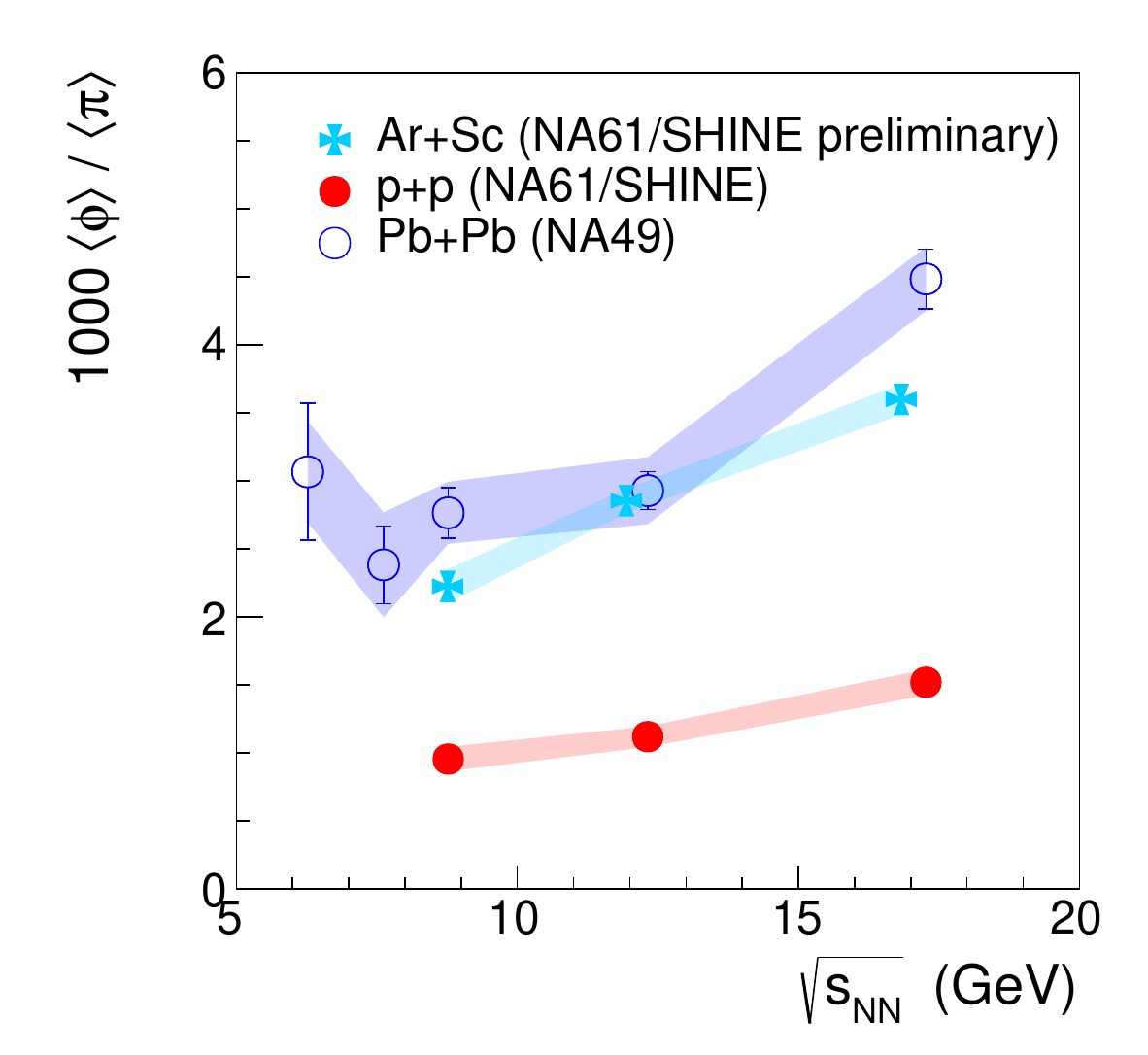}%
  \includegraphics[width=0.3\textwidth, page = 2]{yield_ratios.pdf}%
  \includegraphics[width=0.3\textwidth, page = 3]{yield_ratios.pdf}
  \caption[]{Left: energy dependence of the $\phi/\pi$ ratio for the minimum bias
    \pp~\capcite{NA61SHINE:2019gqe}, central \ArSc, and central
    \PbPb~\capcite{NA49:2008goy} collisions. Middle and right: energy dependence of
    enhancement of \phi and charged kaon production relative to pion production
    in \PbPb and \ArSc collisions
    over the \pp reactions. Yields of pions and kaons
    come from \caprecite{NA61SHINE:2017fne} for \pp collisions, from
    \caprecite{NA61SHINE:2023epu} for \ArSc and from \caprecite{NA49:2002pzu} for
    \PbPb reactions. Vertical bars denote
    statistical uncertainties, while shaded bands represent systematic
    uncertainties.
  }
  \label{fig:yields}
\end{figure*}

\section{Results}
Results were obtained for three \ArSc collision energies: \sNNn{8.8},
\sNNn{11.9}, and \sNNn{16.8}. \Cref{fig:pt} shows double-differential (\y, \pt ) spectra
of \phiM produced in the 10\% most central \ArSc collisions at
\sNNn{16.8}. Similar distributions are also available for the other two
collision energies. The \pt spectra were fitted for each \y bin with thermally
motivated functions:
\begin{equation}
  f(\pt) \propto \pt \cdot \exp(-\frac{\mt}{T})\,,
  \label{eq:pt}
\end{equation}
in order to obtain the integrals of the unmeasured tails of the \pt spectra to calculate
\pt-integrated $\dv*{n}{\y}$ distributions. These tail contributions were small,
not larger than 5\%.
\par
Not constituting an independent result, but rather a different representation of
results described in the previous paragraph, \cref{fig:mt} shows transverse mass
spectra of \phiM mesons produced at midrapidity at three \ArSc collision
energies. The data points were fitted with exponentials to obtain inverse slope
parameters of $T=200 \pm 13 \pm 16$\,\si{\MeV},
$T=226 \pm 12 \pm 22$\,\si{\MeV} and
$T=246 \pm 12 \pm 8$\,\si{\MeV} for respectively \sNNn{8.8},
\sNNn{11.9}, and \sNNn{16.8}. These turn out to be comparable to those for
charged kaons produced in the same collisions~\cite{NA61SHINE:2023epu}, and larger than for \phiM mesons
measured in \pp reactions (about \SI{150}{\MeV})~\cite{NA61SHINE:2019gqe}.
\par
By summing up the measured parts of the double-differential spectra and adding
small tail contributions estimated with fits, rapidity distributions were
obtained. In \cref{fig:dndy} these are compared for the three collisions
energies to model predictions from two variants of \UrQMD~\cite{bib:UrQMD1998,
bib:UrQMD1999} and \Pythia Angantyr~\cite{Bierlich:2018xfw}. One can see that
none of the considered models even comes close to describing the data points,
with both the base variant of \UrQMD and Angantyr significantly underestimating
the data, and \UrQMD with hydrodynamics significantly overestimating the \phiM
production in \ArSc collisions at CERN SPS energies.
\par
Finally, by summing up the measured parts of the rapidity distributions, adding
small (up to 2.5\%) tail contributions estimated with double Gaussian fits, and
doubling thanks to the approximate backward-forward symmetry of central \ArSc
collisions, $4\pi$ \phiM yields were obtained for the three collision energies.
To remove trivial effects of enlarged collision systems, these were divided by
the pion yields calculated as $\expval{\pi}=\frac{3}{2}\pqty{\expval{\pip} +
\expval{\pim}}$~\cite{NA61SHINE:2019gqe}. The energy dependence of the resulting
$\phi/\pi$ ratios for central \ArSc collisions is compared in \cref{fig:yields}
to those for the minimum bias \pp~\cite{NA61SHINE:2019gqe} and central
\PbPb~\cite{NA49:2008goy} reactions. One can see that the $\phi/\pi$ ratio for
\ArSc is  comparable to that in \PbPb collisions, where the onset of
deconfinement is expected, and much higher than for \pp interactions. Next,
double ratios showing enhancement of $\phi/\pi$ over the \pp collisions are
built for \ArSc and \PbPb collisions and compared to similar double ratios for
charged kaons calculated with data from \recites{NA49:2002pzu,
NA61SHINE:2017fne, NA61SHINE:2023epu}. It is clear, that the \phiM enhancement
over \pp collisions is slightly higher than for charged kaons in both \ArSc and
\PbPb reactions, and independent of the collision energy in the considered
range.

\section{Summary}
Preliminary results on the \phiM meson production in the 10\% most central \ArSc
collisions at \sNNn{8.8}, \sNNn{11.9}, and \sNNn{16.8} measured by the
\NASixtyOne experiment were presented. They were obtained using the
tag-and-probe method, utilizing the $\phi \to \Kp\Km$ decay channel. These are
the first-ever results on hidden strangeness production in intermediate-size
systems at CERN SPS. The results include double-differential rapidity and
transverse momentum spectra of \phiM mesons, as well as \pt-integrated rapidity
distributions, transverse mass spectra at midrapidity and $4\pi$ yields. The
rapidity distributions were compared to three microscopic models, which failed
to reproduce the experimental results. Ratios of \phiM to pion production were
considered, showing similarity of values for \ArSc and \PbPb systems, much
higher than for \pp collisions. Finally, hidden and open strangeness production
enhancement was discussed, showing that for \phiM it is slightly higher than for
charged kaons in both \ArSc and \PbPb reactions, and independent of the
collision energy in the considered range.

\vskip3mm \textit{This work was supported by the National Science Centre,
Poland (grant number 2023\slash 51\slash D\slash ST2\slash 02950).}

\newcommand{\Received}{xx.xx.xx}
\bibliographystyle{my-ukr-j-phys}
\bibliography{bibliography}

@article{NA61SHINE:2019gqe,
    author = "Aduszkiewicz, A. and others",
    collaboration = "NA61/SHINE",
    title = "{Measurement of $\phi $ meson production in $p + p$ interactions at 40, 80 and $158 \, \hbox {GeV}/c$ with the NA61/SHINE spectrometer at the CERN SPS}",
    eprint = "1908.04601",
    archivePrefix = "arXiv",
    primaryClass = "nucl-ex",
    reportNumber = "FERMILAB-PUB-19-423-AD-ND-SCD",
    doi = "10.1140/epjc/s10052-020-7675-6",
    journal = "Eur. Phys. J. C",
    volume = "80",
    number = "3",
    pages = "199",
    year = "2020"
}

@article{bib:NA61_facility,
  author = "N. Abgrall and others",
  collaboration = "\NASixtyOne",
  title = "{\NASixtyOne} facility at the {CERN SPS}: beams and detector system",
  journal = "JINST",
  year = "2014",
  volume = "9",
  pages = "P06005"
}

@article{NA61SHINE:2023epu,
    author = "Adhikary, H. and others",
    collaboration = "NA61/SHINE",
    title = "{Measurements of $\pi ^\pm $, $K^\pm $, p and $\bar{p}$ spectra in $^{40}\hbox {Ar+}^{45}\hbox {Sc}$ collisions at 13$A$ to 150$A$~GeV/$c$}",
    eprint = "2308.16683",
    archivePrefix = "arXiv",
    primaryClass = "nucl-ex",
    reportNumber = "CERN-EP-2023-179, FERMILAB-PUB-23-563-AD",
    doi = "10.1140/epjc/s10052-024-12602-2",
    journal = "Eur. Phys. J. C",
    volume = "84",
    number = "4",
    pages = "416",
    year = "2024"
}

@article{NA61SHINE:2017fne,
    author = "Aduszkiewicz, A. and others",
    collaboration = "NA61/SHINE",
    title = "{Measurements of $\pi ^\pm $ , K$^\pm $ , p and ${\bar{\text {p}}}$ spectra in proton-proton interactions at 20, 31, 40, 80 and 158  $\text{ GeV}/c$ with the NA61/SHINE spectrometer at the CERN SPS}",
    eprint = "1705.02467",
    archivePrefix = "arXiv",
    primaryClass = "nucl-ex",
    reportNumber = "CERN-EP-2017-066",
    doi = "10.1140/epjc/s10052-017-5260-4",
    journal = "Eur. Phys. J. C",
    volume = "77",
    number = "10",
    pages = "671",
    year = "2017"
}

@article{NA49:2008goy,
    author = "Alt, C. and others",
    collaboration = "NA49",
    title = "{Energy dependence of $\phi$ meson production in central Pb+Pb
      collisions at $\sqrt{s_{NN}}$ = 6 to 17 GeV}",
    eprint = "0806.1937",
    archivePrefix = "arXiv",
    primaryClass = "nucl-ex",
    doi = "10.1103/PhysRevC.78.044907",
    journal = "Phys. Rev. C",
    volume = "78",
    pages = "044907",
    year = "2008"
}

@article{NA49:2002pzu,
    author = "Afanasiev, S. V. and others",
    collaboration = "NA49",
    title = "{Energy dependence of pion and kaon production in central Pb+Pb collisions}",
    eprint = "nucl-ex/0205002",
    archivePrefix = "arXiv",
    doi = "10.1103/PhysRevC.66.054902",
    journal = "Phys. Rev. C",
    volume = "66",
    pages = "054902",
    year = "2002"
}

@article{NA49:2000jee,
  author = "Afanasiev, S. V. and others",
  collaboration = "NA49",
  title = "{Production of $\phi$-mesons in p+p, p+Pb and central Pb+Pb collisions at $E_\mathrm{beam}=158A$\,GeV}",
  doi = "10.1016/S0370-2693(00)01023-6",
  journal = "Phys. Lett.~B",
  year = "2000",
  volume = "491",
  pages = "59"
}

@article{ATLAS:2014pju,
    author = "Aad, Georges and others",
    collaboration = "ATLAS",
    title = "{The differential production cross section of the $\phi $ (1020) meson in $\sqrt{s}$ = 7 TeV $pp$ collisions measured with the ATLAS detector}",
    eprint = "1402.6162",
    archivePrefix = "arXiv",
    primaryClass = "hep-ex",
    reportNumber = "CERN-PH-EP-2012-269",
    doi = "10.1140/epjc/s10052-014-2895-2",
    journal = "Eur. Phys. J. C",
    volume = "74",
    number = "7",
    pages = "2895",
    year = "2014"
}

@article{LHCb:2011ijs,
    author = "Aaij, R. and others",
    collaboration = "LHCb",
    title = "{Measurement of the inclusive $\phi$ cross-section in $pp$ collisions at $\sqrt{s}=$ 7 TeV}",
    eprint = "1107.3935",
    archivePrefix = "arXiv",
    primaryClass = "hep-ex",
    reportNumber = "LHCB-PAPER-2011-007, CERN-PH-EP-2011-106",
    doi = "10.1016/j.physletb.2011.08.017",
    journal = "Phys. Lett. B",
    volume = "703",
    pages = "267",
    year = "2011"
}

@article{Gazdzicki:2010iv,
    author = "Gazdzicki, M. and Gorenstein, M. and Seyboth, P.",
    title = "{Onset of deconfinement in nucleus-nucleus collisions: Review for pedestrians and experts}",
    eprint = "1006.1765",
    archivePrefix = "arXiv",
    primaryClass = "hep-ph",
    doi = "10.5506/APhysPolB.42.307",
    journal = "Acta Phys. Polon. B",
    volume = "42",
    pages = "307",
    year = "2011"
}

@article{bib:UrQMD1998,
  author = "S.A. Bass and others",
  title = "Microscopic models for ultrarelativistic heavy ion collisions",
  journal = "Prog. Part. Nucl. Phys.",
  year = "1998",
  volume = "41",
  pages = "255"
}

@article{bib:UrQMD1999,
  author = "M. Bleicher and others",
  title = "Relativistic hadron-hadron collisions in the ultra-relativistic quantum molecular dynamics model",
  journal = "J. Phys.~G",
  year = "1999",
  volume = "25",
  pages = "1859"
}

@article{Bierlich:2018xfw,
    author = {Bierlich, Christian and Gustafson, G{\"o}sta and L{\"o}nnblad, Leif and Shah, Harsh},
    title = "{The Angantyr model for heavy-ion collisions in PYTHIA8}",
    eprint = "1806.10820",
    archivePrefix = "arXiv",
    primaryClass = "hep-ph",
    reportNumber = "LU-TP-18-19, LU-TP 18-19, MCnet-18-12",
    doi = "10.1007/JHEP10(2018)134",
    journal = "JHEP",
    volume = "10",
    pages = "134",
    year = "2018"
}


\vspace*{-5mm} \rezume{%
Authors in Ukraine. Don't fill it.} {Title in Ukrainian. Don't fill it. } {Abstract in Ukraine. Don't fill it. The editorial board will fill it correctly in Ukraine.
}{\textit{К\,л\,ю\,ч\,о\,в\,і\,
с\,л\,о\,в\,а:} keywords in Ukraine. Don't fill it.}
\end{document}